\documentclass{iauJDSS}
\usepackage{graphicx}

\title[Rapid Mass Segregation] 
{Rapid Mass Segregation in Massive Star Clusters}

\author[McMillan et al.]   
{Stephen McMillan$^1$, Enrico Vesperini$^2$, \and Nicholas Kruczek$^1$}

\affiliation{$^1$Department of Physics, Drexel University,
  Philadelphia, PA 19104, USA \break Contact email: steve@physics.drexel.edu
  \\[\affilskip]
  $^2$Department of Astronomy, Indiana University, Swain Hall West,
  Bloomington, IN 47405, USA}

\pubyear{2012}
\volume{Volume xx}  
\pagerange{119--126}
\date{}
\jname{Highlights of Astronomy, Volume 14}
\editors{Giampaolo Piotto and Enrico Vesperini, eds.}

\begin{document}

\maketitle
\begin{abstract}
  Several dynamical scenarios have been proposed that can lead to
  prompt mass segregation on the crossing time scale of a young
  cluster.  They generally rely on cool and/or clumpy initial
  conditions, and are most relevant to small systems.  As a
  counterpoint, we present a novel dynamical mechanism that can
  operate in relatively large, homogeneous, cool or cold systems.
  This mechanism may be important in understanding the assembly of
  large mass-segregated clusters from smaller clumps.
\end{abstract}


Early mass segregation may be critical to the long-term survival of a
stellar system (Vesperini et al. 2009a). 
It also defines the early cluster environment within which stars move
and interact.  In recent years, several dynamical studies have
explored routes to early mass segregation that do not simply require
that a cluster formed in that state.  \cite{McMillan_etal2007} found
that mergers of mass-segregated ``clumps'' tend to preserve that
segregation in the final merger product, so that, if the clumps are
formed segregated, or have time to segregate before they merge, the
result is a strongly mass-segregated cluster.  \cite{Allison_etal2009}
found similar behavior, starting from fractal clumpy initial
conditions in small, cool model clusters.

\begin{figure}[b!]
  ~~
  \includegraphics[width=3in, bb = 30 20 680 490,
	clip = true]{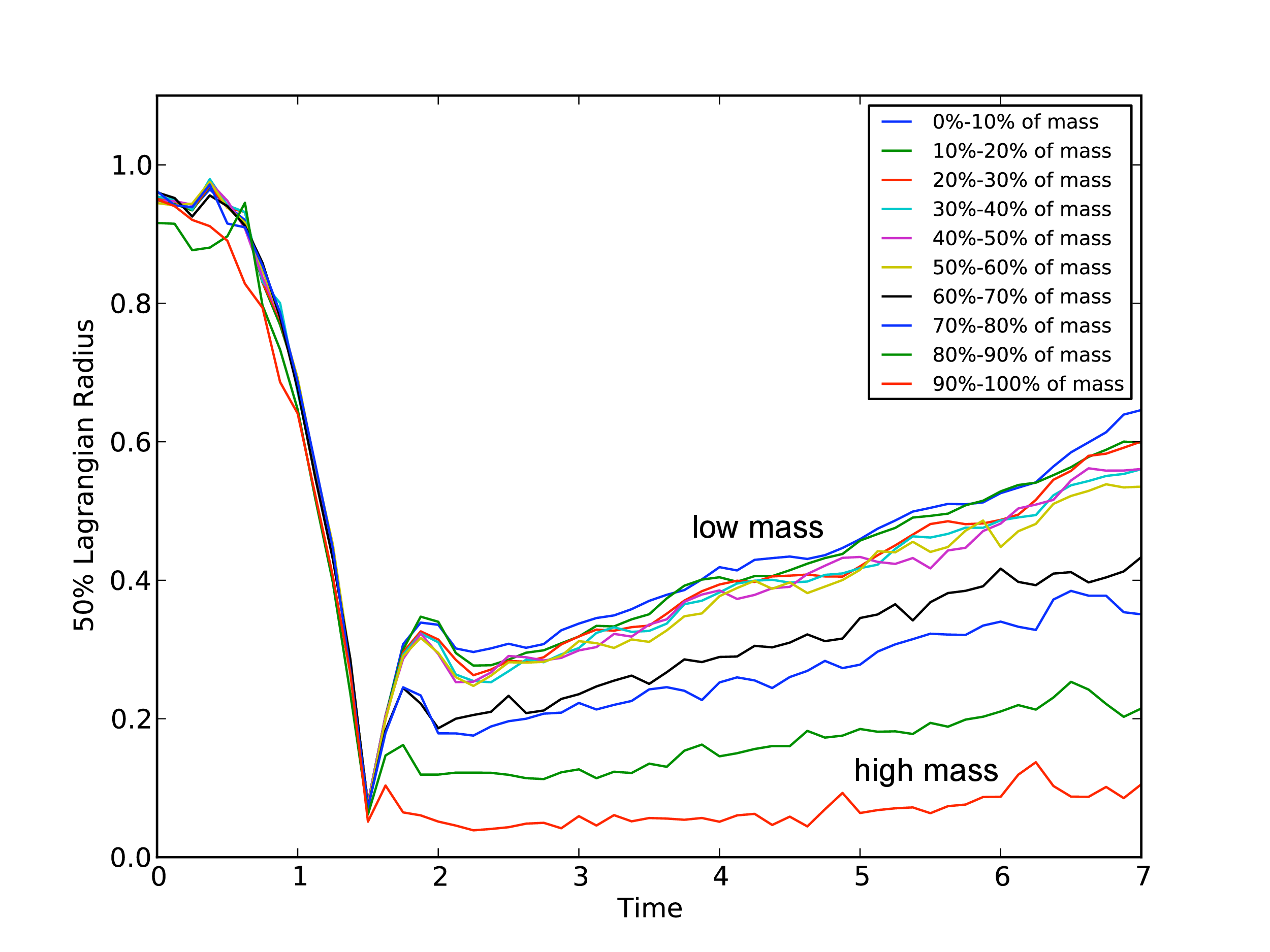}
  ~~
  \parbox[b]{2in}{{\bf Figure 1.} Collapse of an initially cold,
    homogeneous spherical system containing $10^4$ particles with a
    Kroupa (2002) mass function.  The half-mass radii of the particles
    making up the bottom 10 percent, 10--20 percent, 20--30 percent,
    etc., of the cumulative mass distribution are shown.  The bottom
    four lines after the collapse represent the top four mass groups,
    their half-mass radii decreasing with mass, indicating strong mass
    segregation.\\~}
\end{figure}

Ultimately, these scenarios rely on normal relaxation processes in
small stellar systems.  However, as illustrated in Figure 1, rapid
segregation is also possible in significantly larger systems.  The
initial conditions of the simulation shown here consist of a cold
(virial ratio $q=0.001$), homogeneous sphere with a \cite{Kroupa2002}
mass distribution.  No segregation is seen before the ``bounce'' at
$t\sim1.5$ initial dynamical times, while immediately afterward the
highest mass groups are clearly ordered by radius.  This behavior was
also noted by \cite{Vesperini_etal2006} and
\cite{Vesperini_etal2009b}.

The phenomenon of rapid segregation cannot be due to enhanced
relaxation around the high density bounce.  This would only be
possible if the system were still cold at that time, and our
simulations clearly indicate that this is not the case.  Instead, as
shown in Figure 2, we find that the system fragments as it collapses,
as discussed in detail by \cite{Aarseth_etal1988}, and the fragments
mass segregate quite early on during the collapse process.
Significant segregation within the clumps is already established by
$t\sim1$, well before the bounce, and is preserved when the clumps
subsequently merge at $t=1.5$, essentially as described by
\cite{McMillan_etal2007}.

\begin{figure}[t!]
  ~~
  \includegraphics[width=3in, bb = 100 100 1330 1330,
	clip = true]{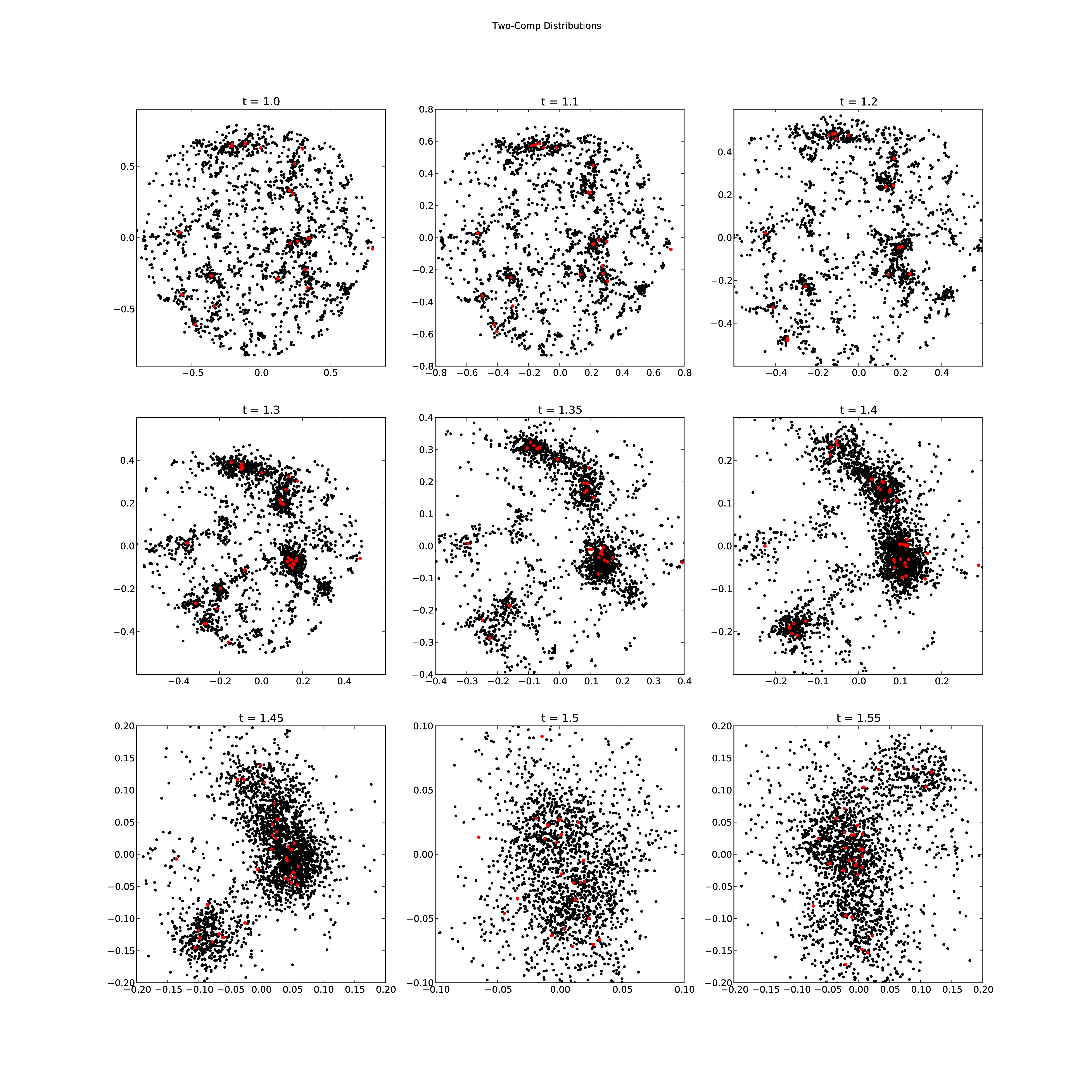}
  ~~~~
  \parbox[b]{1.75in}{{\bf Figure 2.} Fragmentation and mass
    segregation during the collapse can be seen in this sequence of
    frames running from $t=1$ to $t=1.55$, just after the moment of
    collapse. The spatial scale of the frames shrinks to follow the
    collapsing system, from $\pm0.9$ at top left ($t=1$) to $\pm0.4$
    at center ($t=1.35$), to $\pm0.1$ at bottom center ($t=1.5$) and
    $\pm0.2$ at bottom right ($t=1.55$).\\~}
\end{figure}

The phenomenon persists as we vary the initial system parameters, and
is still measurable even for fairly ``warm'' initial conditions
($q\sim0.1$), with and without initial clumping (fractal dimension
$d\sim2-3$), and for large systems, up to $N\sim10^5$.  Thus it may
provide the basis of a viable mechanism for extending earlier
dynamical segregation scenarios to substantially larger systems.

\begin{acknowledgments}
  This work was supported in part by NSF grants AST-0708299 and
  AST-0959884, and NASA grant NNX08AH15G.
\end{acknowledgments}

\end{document}